# NLP-SIR: A Natural Language Approach for Spreadsheet Information Retrieval

Derek Flood, Kevin Mc Daid, Fergal Mc Caffery

Dundalk Institute of Technology

derek.flood@dkit.ie, kevin.mcdaid@dkit.ie, fergal.mccaffery@dkit.ie

**ABSTRACT**

*Spreadsheets are a ubiquitous software tool, used for a wide variety of tasks such as financial modelling, statistical analysis and inventory management. Extracting meaningful information from such data can be a difficult task, especially for novice users unfamiliar with the advanced data processing features of many spreadsheet applications. We believe that through the use of Natural Language Processing (NLP) techniques this task can be made considerably easier. This paper introduces NLP-SIR, a Natural language interface for spreadsheet information retrieval. The results of a recent evaluation which compared NLP-SIR with existing Information retrieval tools are also outlined. This evaluation has shown that NLP-SIR is a more effective method of spreadsheet information retrieval.*

## 1. INTRODUCTION

The latest version of Microsoft© Excel can contain over 1 million rows per worksheet facilitating the storage of considerable amounts of data. With such a large volume of data, extracting meaningful information can be a difficult task. Features such as Filters and PivotTables can assist. However, for novice spreadsheet users the use of these features can cause great difficulty.

The aim of this research is to investigate and improve the effectiveness and efficiency of spreadsheet information retrieval. One way this improvement can be achieved is through a Natural Language Interface. Natural Language Interfaces, or NLI's, allow a user to operate a computer application by telling the computer what they want, in their own way. The application then interprets this, through the use of NLP techniques, and performs the desired action on behalf of the user.

NLI's have been demonstrated to improve the efficiency with which users interact with applications. Chart generation [Kato, Matsushita et al., 2002] and accessing databases [Liddy and Liddy, 2001] are just some of the domains which have benefited from the use of NLI's. There are many ways in which NLP may be performed, some of which are outlined in section 2.

NLP-SIR is a NLI for spreadsheet information retrieval. The system allows users to perform common information retrieval tasks, such as filtering and generating summary tables, similar to PivotTables, through the use of natural language. Section 3 details the NLP-SIR system and also highlights some of the limitations of the current implementation of the system.

The effectiveness and efficiency of the NLP-SIR system were evaluated through a controlled experiment. This experiment compared NLP-SIR with existing information retrieval tools found within Microsoft© Excel. A full description of this evaluation along with the results can be found in Section 4. Section 5 outlines some ways in which this technology can be improved. Section 6 then concludes this paper

**2. BACKGROUND**

Spreadsheets are a versatile software application. They allow complex simulations to be run quickly and different scenarios to be modelled with little modification. All of these tasks are accomplished through the use of formulae, however a large number of spreadsheets contain no formulae at all. These spreadsheets are used purely to store data which can be later analysed for additional information.

The Euses spreadsheet corpus[Fisher and Rothermel, 2005] is a collection of over 4,000 real-world spreadsheets, which have been collected from a variety of sources primarily the World Wide Web. An inspection of this corpus, by the author, has shown that the three largest spreadsheets, each containing over 1 million cells, contain no formulae. These spreadsheets instead contain static data which can be interrogated for more detailed information.

Spreadsheet tools such as PivotTables and Filters can be used to help analyse this data, however for novice spreadsheet users, these features can be difficult to use effectively. Although more experienced users find using these features to be straightforward, they still require multiple interactions, which can be time consuming.

There are many ways in which complex features like these can be simplified. One approach is the use of a command based interface. In previous work [Flood and McDaid, 2007] the authors have developed a command based navigation system to simplify the way in which users of voice recognition technology navigate a worksheet. This approach allowed the user to perform certain navigational tasks, which would have taken multiple interactions with existing voice recognition technology, more efficiently.

Another approach that may simplify these tasks is through NLP. NLP is a way for computer applications to interpret human language. By using a NLI a user can operate a computer by telling the computer what they want in their own way. The input from these interactions is often referred to as an utterance. A number of applications[Zelle and Mooney R. J., 1996; Liddy and Liddy, 2001; Tang and Mooney, 2001; Kato, Matsushita et al., 2002; Begel, 2005; Woodley, Tannier et al., 2006], including chart generation, have benefitted from NLI's

A common way to analyse large amounts of data is through charts. Charts can also be used to highlight characteristics of data that would be otherwise indistinguishable. As these characteristics become visible, they will lead the user to ask further questions which can then be answered through additional charts. Kato et al [Kato, Matsushita et al., 2002] propose the use of a natural language interface to help users in this type of exploratory data analysis.

Kato's system allows a user to express, in natural language, what they would like to have presented in a chart. When a chart has been generated the user can then alter it through additional utterances to further interrogate the data. For example, if a user was looking at sales data for Ireland over the last ten years and wanted to see a visual representation of just those sales made in County Louth in 2004 and 2005, they might say "Show me the sales in Louth for 2004 and 2005". The system would then respond with a chart showing the sales in Louth for those 2 years. After seeing this chart the user may wish to see this same information broken down by town and might say "By town", which would cause the system to alter the chart to show the sales in Louth for 2004 and 2005 broken down by town.

In using this approach, the user does not need to go through the steps of manually creating each chart. Instead the system can automatically determine the right type of chart and the correct data to include before generating the correct chart. The system can also determine changes to the parameters when subsequent utterances are given. This approach has the added benefit of enabling inexperienced users to generate charts quickly and with minimum effort.

In order to evaluate the effectiveness of the system, Kato et al[Mitsunori, Eisaku et al., 2004], asked 25 participants to select the chart, from multiple possibilities, which most accurately depicted the information required by a natural language statement. The participants and the system were presented with six different statements and in 4 of these scenarios participants selected the same chart as the system generated. In the remaining cases the chart generated by the system was the second most popular.

Reducing the number of interactions is not the only way in which NLI's can improve users' effectiveness. Structured Query Language (SQL) may be used by users to retrieve information from a database. In order to use SQL effectively however users need to know the structure of the database as well as the syntax of the SQL language. This scenario is not ideal for users who have no background in database technologies.

In order to facilitate the processing of queries from users with no statistical background but who needed access to statistical information, Liddy et al[Liddy and Liddy, 2001] have developed a natural language based system. This system uses NLP techniques to interpret the information that is being sought from a natural language style query. By testing the system on a small sample of user queries, Liddy et al found that 95% of the new user queries could be covered and covered accurately by their system.

3. **NLP-SIR**

3.1. **System Overview**

When the concept of a natural language interface to spreadsheets was first conceived a number of characteristics of such a system were immediately apparent. As a wide variety of spreadsheets are used in different situations it became evident that a NLI for spreadsheet information retrieval must be adaptable and could not be tied to any one domain. A number of NLI's, especially those in the database domain, for example[Zelle and Mooney R. J., 1996], are constructed for a particular database. Therefore, applying the same system to another domain would not be successful without some modification or additional training.

Another characteristic that is essential to a NLI for spreadsheet information retrieval is flexibility. In order to accommodate natural language any system should be able to identify the meaning of an utterance independent of the structure of that utterance. For example consider a spreadsheet containing a list of golf courses situated in the state of Indiana. Such a spreadsheet could contain information such as the number of holes on the course, the type of terrain and the price of playing on the course. While looking at this spreadsheet, one user might say "*what golf courses in Marion have executive difficulty*" Where as another user might ask for a "*list of golf courses that are executive and in Marion*". Both users are looking for the same information but the way in which they ask is structured differently.

The result of incorporating these characteristics into a NLI for spreadsheets is NLP-SIR, (Natural Language Processing for Spreadsheet Information Retrieval). This system allows a user to perform certain information extraction functions by telling the system what they would like in their own way. The present system allows the user to filter out rows or columns, to count the number of rows that meet a certain set of criteria and to find the most or least frequent value in a given column. It also allows users to generate tables, similar to PivotTables which count the number of rows that meet certain criteria.

The current implementation of the system uses a text based interface where a user types in what they would like to ask the system. The answer generated by the system is presented to the user through an alert box, similar to that shown in Figure 1.

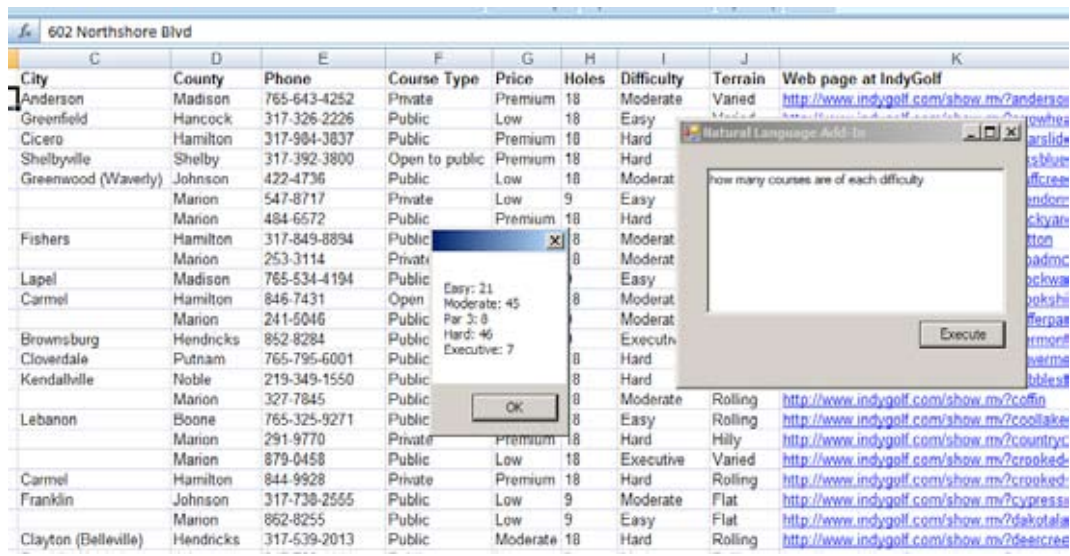

**Figure 1: NLP-SIR**

The spreadsheet depicted in Figure 1 shows a list of golf courses that are situated in the state of Indiana. There are over 120 courses in total and extracting meaningful information, such as the number of courses of each difficulty, from this data can be time consuming. To find this information using the conventional interface, a user would need to create a PivotTable doing a count of each difficulty type. With NLP-SIR the same user could type in the question "*How many courses are of each difficulty?*" and the system would find the information and present it to them, eliminating the need for them to manually create a PivotTable.

During the course of developing NLP-SIR, a number of natural language systems were evaluated to see if any of these existing approaches could be used within the spreadsheet domain. Domains such as databases [Zelle and Mooney R. J., 1996], Chart generation [Kato, Matsushita et al., 2002] and XML retrieval [Woodley, Tannier et al., 2006] were all examined and each system offered a unique approach to the problem. It was found however that none of these systems contained all of the desired characteristics. The

           

approach used by NLP-SIR to infer meaning from the utterance has similarities to the systems presented above.

### 3.2. System Assumptions and Limitations

The present implementation makes certain assumptions about the spreadsheet. It assumes that the data is well structured and that each row in the spreadsheet corresponds to one entry. The system also assumes that the first row within the spreadsheet contains the name of each column. As well as this it is assumed that only one data type is featured within each column. The spreadsheet structure supported by NLP-SIR is a common structure for storing data.

In addition to the above assumptions NLP-SIR is limited in the utterances it will correctly recognise. When users are referring to a column, they must use the column name as it appears in the first row of the spreadsheet. For example if a user wanted to refer to the Price column in the above spreadsheet they would need to use the word "Price" and not other synonyms, such as cost. In certain instances whitespace would appear within column headings. The NLP-SIR system ignores whitespace and looks only at the characters that appear within the heading.

When a user wants to refer to a numeric column the user must explicitly name this column. With textual columns the user can use the text as it appears within the spreadsheet without the need to state the name of the column which contains the text. For example if a user wanted to see all of the courses with an easy difficulty they could say "*Show me all of the easy courses*". The system can automatically recognise the column in which the value "easy" appears.

If a value such as this appears in many columns, the system can ask the user for clarification before applying the desired function. The system will alert the user to the columns the data element appears in while asking for clarification. This way the user can see exactly what the system did not understand.

The present implementation of the system is based on positive statements, i.e. the user instructs the system as to what they want. In some circumstances it may be desirable to tell the system what they don't want, for example if a user wanted to see all of the golf courses that did not have a varied terrain. This type of query is not implemented in the current system; however this will be included in future versions of NLP-SIR.

Another limitation of the system is the output mode. When the system displays additional information, such as pivot tables they are displayed in an alert box. This is not an ideal approach as it causes users difficulty in evaluating the systems results. If the results are quite large then the alert box exceeds the screen size and the user is unable to see part of the results.

### 4. EVALUATION

In order to test the effectiveness of the NLP-SIR system, a comparative study was conducted. The initial evaluation focused on novice spreadsheet users as it is believed that this group of users would benefit most from this new technology. It is hypothesised that novice spreadsheet users would be able to perform information retrieval tasks, more effectively and more efficiently using NLP-SIR than through using the existing Microsoft© Excel interface.

## 4.1. Experiment design

In order to test this hypothesis, 41 participants were asked to take part in a comparative experiment. The participants selected were 4th year students studying for a Bachelor of Business Studies (Honours). One of the modules that the participants had taken was on spreadsheets and in particular Microsoft© Excel. Although the participants had used Microsoft© Excel in previous years, it is believed that, for most participants, this year was the first time that they used the 2007 version including the updated interface.

As part of their spreadsheet course, the participants were divided into four groups in order to ensure adequate facilities were available for each student. Each group had a one hour lab class per week and it was during these lab classes that the experiment was conducted. All participants were given a brief description of the study.

Once the participants had agreed to take part, they were given a brief overview of the experiment. Each class group was randomly divided into two groups, Group 1 and Group 2. Group 1 were asked to complete the tasks using the existing technology while group 2 were asked to use the NLP-SIR system. Before they began the task, the participants were given a brief recap on how to use Filters and PivotTables on a sample spreadsheet. They were also shown how to perform the same tasks using the NLP-SIR system.

For the evaluation, participants were asked to retrieve information from a given spreadsheet. This spreadsheet contained a list of 127 golf courses within the state of Indiana. Information such as the type of terrain, the number of holes and the difficulty of each course was detailed in this spreadsheet. The participants were told that they were employed in a tourist office and were asked to write down the information that would satisfy some requests for information that had been received. These requests, similar to R1, asked for either a list of the courses that met a certain set of criteria or the number of courses that met these criteria.

> **R1:** Hi, I am a novice golfer and I am going on holidays to Hancock. While there I would like to try some of the easy courses. Can you provide me with a list of such courses with a varied terrain?

The requests for information were displayed to the participants in a task pane similar to that in Figure 2. When the participant had found the information they needed and written it down, they were required to press the "next task" button to move on to the next task. When they had completed the last task a "thank you" message was displayed.

In total 10 tasks were given to each participant. These tasks were divided into three levels of difficulty; Easy, Intermediate and Complex. The tasks were divided based on how difficult it would be to accomplish them using the existing technology. The first 4 tasks were classified as easy and required participants to filter the data, based on a set of criteria, such as having a terrain of flat and a difficulty of easy, or to find the most common type of terrain. The next 3 tasks were classified as intermediate and asked participants to create a PivotTable with a varying number of filters and columns. One of these tasks asked participants to find the number of executive courses in each county. The final 3 tasks were classified as complex and required participants to perform an "or" type query over multiple columns. One such task asked the participants to find the number of courses that either had a hilly terrain or had a difficulty level of hard.

**Figure 2: The Task Pane**

No time limit was set for the task, allowing users to finish when they felt they had completed all of the tasks. The majority of participants had attempted all of the tasks however some participants only attempted the first few tasks. When the participants had finished the experiment they were asked to complete a short questionnaire regarding their experience of the technology and of the tasks. The aim of this questionnaire was to evaluate the perceived experience of using the technologies in a given situation.

### 4.2. Results

While the participants were doing the task, a customised macro was run in the background to record the time of each cell change and each cell selection, the time and details of each filter that was applied to the spreadsheet and finally the utterances that were given to and results that were received from the NLP-SIR system. This information was used as a basis for evaluating the effectiveness and efficiency of the system.

The effectiveness of the system was measured using the number of tasks that the participants had completed successfully and the number of utterances that were required for each task. The efficiency of the system was evaluated using the total time that it had taken participants to complete each task.

The recording macro failed to record data for 3 participants resulting in data for just 38 participants. 20 of these participants had used Microsoft© Excel for the task while the remaining 18 had used the NLP-SIR system.

**Successfully Completed Tasks**

During the experiment participants were asked to write down the information that would answer the requests that they had been given. The first three tasks required users to write down the name of the golf courses that would satisfy a given set of criteria. The fourth task then asked participants for the most popular type of terrain. The remaining tasks asked users to write down the number of courses that met a certain set of criteria.

Each participant's written answers were manually reviewed by the author to determine if the answers they supplied matched the pre-determined answers. In some cases the authors' judgement was used to determine if the supplied answer was in fact correct, for

Copyright © 2009 EuSpRIG & The Author(s)    7

example some participants wrote down the address of the course instead of the name, in this particular scenario the answer was deemed to be correct. Only a small number of exceptional cases arose and were seen in both sets of results. In each case the judgements made were consistent.

| | |
|---|---|
| **Excel** | 3.43 |
| **NLP-SIR** | 6.75 |
| **Total Number of Tasks** | 10 |

**Table 1: Average number of tasks completed successfully**

Table 1 shows the average number of tasks that were successfully completed by participants. On average the participants who were using the NLP-SIR system were able to complete 6.75 of the 10 tasks correctly. The participants who used the Microsoft© Excel technology however were only able to correctly complete, on average, 3.43 of the tasks. Using non-parametric methods the statistical significance of these results were measured and a p-value of .0003 was found indicating that the difference in performance is statistically significant. In order to investigate further where the NLP-SIR system was of most benefit to participants, the number of tasks completed successfully in each category was also examined.

| | **Easy** | **Intermediate** | **Complex** |
|---|---|---|---|
| **Excel** | 2.85 | 1.10 | 0.05 |
| **NLP-SIR** | 3.22 | 1.06 | 1.78 |
| **P-Value** | .13* | - | .00* |
| **Total Number of Tasks** | 4 | 3 | 3 |

**Table 2: Average number of tasks completed successfully by category (*P-value given to 2 places of decimal)**

Table 2 shows the average number of tasks that were completed successfully for each category of task. This table shows that participants who were using the NLP-SIR technology performed better while doing the Easy and Complex tasks; however it was found that the opposite occurred with the intermediate tasks.

In order to determine why this was the case, the utterances that were used by the participants of the NLP-SIR system were examined. It was observed that some of the utterances that were given to the system used phrases, such as "each of the counties", that were given in the scenario description. In the case of Task 5 and Task 7, both of which are of intermediate difficulty, these phrases were not recognised by the system and the answer that was given to the participants was incorrect. It is planned to investigate this issue further to determine the extent with which the phraseology of the scenarios influenced the participants' choice of language.

It should also be noted that only one of the participants using Microsoft© Excel was able to complete any of the complex tasks. This participant correctly answered one of these tasks while the participants who had used the NLP-SIR system were able to successfully complete on average 1.78.

**Average time to complete each task**

The time at which each user clicked on the "Next Task" button was recorded during the experiment. This time was used to determine how long each participant spent on each task. The task time was calculated by taking the time each participant started the task away from the time they moved to the next task. If the task pane was closed during the trial, participants would have had to return to the beginning and move through each of tasks again until they reached the task they were on prior to its closure.

While the authors were introducing the technology, some of the participants opened the spreadsheet and clicked onto the first task. They then waited until instructed before beginning the task. This has meant that in some cases the recorded time was not a true reflection of how long it took to complete the first task as it included the time spent by participants listening to the instructions that were being given.

When participants reached the tenth and final task, some of the participants failed to click through to the "thank you" message. This has meant that the time to complete the final task is not a true reflection of the actual time it took the participants to complete the task.

For these reasons the first and final tasks have not been included in the averages presented here. As these two tasks have been discarded, the average times presented here are based on three easy tasks, three intermediate tasks and two complex tasks.

|  | **Easy** | **Intermediate** | **Complex** |
| --- | --- | --- | --- |
| **NLP-SIR** | 136.13 | 118.10 | 97.19 |
| **EXCEL** | 147.45 | 134.05 | 157.28 |
| **Difference** | 11.32 | 15.94 | 60.09 |

Table 3: Average time per task (in seconds)

Table 3 shows the average amount of time that was spend by participants performing each task. It can be seen that for all categories of tasks, the participants who were using NLP-SIR, completed the tasks faster than those using Microsoft© Excel. As the complexity of the task increased, the savings made by participants using the NLP-SIR system also increased with the difference in time on complex tasks being the only statistically significant result.

It can be seen that participants using NLP-SIR spent most time on the easy tasks and less on the subsequent tasks. One reason for this is that participants had to write more information for the easier tasks. These tasks required users to write down the names of the golf courses that satisfied the given set of criteria. It is also believed that as these tasks were the first to be attempted, users spent some time learning to use the system during these tasks. Further analysis will be conducted in the future to determine the total time that was spent by participants interacting with the system.

**Average Number of Utterances per task**

During the experiment, participants did not always get the answer they wanted from the system on the first utterance. If this happened the participant would need to rephrase the utterance and try again. For this reason the average number of utterances required to complete each task has been evaluated.

It was found that during the experiment participants using the NLP-SIR system needed, on average, 1.97 utterances to complete each task. The highest number of utterances given to the system by any one participant on a single task was 12. The reason for this was that the participant looked for additional information beyond what was required for the task. Ultimately the correct answer was recorded by the participant.

On a number of occasions it was found that participants asked the same question, in the same way multiple times. This may have been due to the way the results are displayed in an alert box. When the alert box is shown on screen the user is not able to go back to the spreadsheet without first closing it. Therefore if a participant wanted to check the answer against the data before writing it down, they would need to close the alert box, look at the data and determine if the answer was correct and then ask the question again so that it could be displayed on screen while they wrote it down.

**Types of utterances received**

During the experiment participants used many different ways to ask for the same information. It was noticed that a number of participants used "Google" style queries where they only input the least amount of words necessary to express what they wanted, for example "*Boone flat Golf Course*" or "*number of executive courses in each county*". Other participants wrote full statements, as if they would give to another person such as "*How many flat golf courses are in boone?*" or "*How many executive courses are availible in Marion?*". (The previous utterance is presented as it was input by the user to the system, including the misspelling of the word "available").

In some instances the system was not able to interpret the utterances the participants provided. In such cases the system would not perform any actions and the participant would be required to rephrase the utterance. One example of this occurred when a participant was trying to find the most common type of terrain in the spreadsheet. The first utterance given by the user was "*Most used terrain*". This utterance did not produce any results so the user then re-phrased the utterance to "*Most popular terrain*" to which the system responded with the correct answer.

It was also seen that a small number of participants copied the request directly into the NLP-SIR input box. On some occasions this approach produced the correct answer while on others the results produced were incorrect.

**Limitations of the study**

The results presented here are based on a well structured spreadsheet, which conforms to the assumptions that are made by the NLP-SIR system. Despite these limitations the results demonstrate that natural language is an effective means of retrieving information from spreadsheets. Further work on the NLP-SIR system may provide a more general NLI to spreadsheet information retrieval.

It is believed that the participants that took part in the study are representative of the majority of novice spreadsheet users. These participants had only a minimal amount of training in the features that were required for the tasks like the majority of novice spreadsheet users.

## 5. FUTURE WORK

During the experiment a number of suggestions were made as to how the technology could be improved. The present implementation of the system presents users with a separate window in which they can type their questions. During the task however, participants found that the constant switching between the Microsoft© Excel spreadsheet and the NLP-SIR window was annoying. In order to address this, the system could use a task pane similar to the one that was used to present the tasks, thus allowing users to freely and easily move between the Microsoft© Excel spreadsheet and the NLP-SIR interface.

10Copyright © 2009 EuSpRIG & The Author(s)

Although the participants found the system very helpful, they were not satisfied with the way the information was presented. The current system presents answers in an alert box, which needs to be closed before the user can go back to the spreadsheet. It was suggested that results, such as tables, should be placed on a blank area of the spreadsheet instead of the alert box. This approach will allow the user to interrogate the spreadsheet while looking at the results.

In its current implementation NLP-SIR only presents the final value of a count operation to the user. This approach does not allow the user to verify that the utterance has been understood correctly or to see which values have been included in the count. To address this issue an additional feature will be implemented to allow a user to click on the result of such a count operation and have the rows that have contributed to this value to be highlighted. This approach will allow the user to verify that the right criteria have been used in performing the count operation and thus increase users' confidence in the answers produced by the system.

It is believed that some of the limiting assumptions that are made by NLP-SIR could be removed through further analysis of data spreadsheets.

The research to date has focused on novice spreadsheet users. Although these users will benefit most from the technology, they are not the only ones who will benefit. It is our belief that as the complexity of the task increases the benefits of a natural language approach will be seen by more experienced users. In order to validate this claim a similar experiment will be run in the future with expert spreadsheet users.

## 6. CONCLUSIONS

This paper introduces NLP-SIR, a Natural language approach to spreadsheet information retrieval. Natural Language Interfaces, such as NLP-SIR, allow users to operate computer applications through their own language, eliminating the need for users to conform to existing interfaces. Such approaches are more intuitive to users and encounter less resistance from new users.

An evaluation of NLP-SIR was conducted to investigate its effectiveness compared with Microsoft© Excel for information retrieval. This evaluation asked 41 Novice spreadsheet users to retrieve ten pieces of information from a given spreadsheet. Approximately half of the participants used Microsoft© Excel for the task while the remainder used the NLP-SIR system.

It was found that those participants who used the NLP-SIR system were able to successfully complete more of the tasks than those who used the existing application interface. It was also found that the NLP-SIR users were able to complete these tasks in less time.